\documentclass[12pt]{article}

\usepackage{latexsym}

\usepackage{graphicx}

\begin{document}


\title{Black Saturn with dipole ring
}

\author{
     Stoytcho S. Yazadjiev \thanks{E-mail: yazadj@theorie.physik.uni-goettingen.de; yazad@phys.uni-sofia.bg}\\
{\footnotesize  Institut f\"ur Theoretische Physik, Universit\"at G\"ottingen,}\\
{\footnotesize  Friedrich-Hund-Platz 1, D-37077 G\"ottingen, Germany}\\
{\footnotesize and} \\
{\footnotesize  Department of Theoretical Physics,
                Faculty of Physics, Sofia University,}\\
{\footnotesize  5 James Bourchier Boulevard, Sofia~1164, Bulgaria }\\
}

\date{}

\maketitle

\begin{abstract}
We present a new stationary, asymptotically flat  solution of 5D Einstein-Maxwell gravity describing a Saturn-like black object:
a  rotating  black hole surrounded by a rotating dipole black ring. The solution is generated by combining the  vacuum black Saturn solution
and  the  vacuum black ring solution with appropriately chosen parameters. Some basic properties of the solution are analyzed and
the basic quantities are calculated.
\end{abstract}


\sloppy

\section{Introduction}

In recent years the higher dimensional gravity is attracting much interest. Apart from the fact
that the higher dimensional gravity is interesting in its own right, the increasing amount of works
devoted to the study of the higher dimensional spacetimes is inspired from the string theory and
the brane-world scenario with large extra dimensions.
The gravity in higher dimensions exhibits much richer dynamics and spectrum  of
solutions than in four dimensions. One of the most reliable routes for  better understanding of higher
dimensional gravity and the related topics are the exact solutions. For example, recently discovered
exact black rings solutions\footnote{The black ring solution of \cite{ER2} is with one rotational parameter. Black rings with two
independent rotational parameters have been constructed recently in \cite{Pomeransky}.} with unusual horizon topology \cite{ER2}, demonstrated explicitly
that the 5D Einstein gravity
exhibits unexpected features completely absent in four dimensions. It was shown in \cite{ER2} that both
the black hole and the black ring can carry the same conserved charges, the mass and a single angular
momentum, and therefore there is no uniqueness theorem in five dimensions.
Moreover, the black rings can also carry nonconserved charges which can be varied continuously without
altering the conserved charges. This fact leads to
continuous (classical) non-uniqueness \cite{EMP}.

Among of the most fascinating  solutions of the 5D vacuum Einstein gravity  discovered so far are those describing many black objects in equilibrium
 \cite{ElvangFigueras}, \cite{IgichiMishima} (see also \cite{Gauntlett1} and \cite{Gauntlett2} in the case of 5D  supergravity). In particular we will be interested in the so-called black Saturn solution describing a rotating
black hole surrounded by a rotating black ring. The natural question is whether a similar solution exists in the more general case
of non-vacuum 5D gravity, especially in 5D Einstein-Maxwell gravity.
The aim of this paper is to present a new stationary, asymptotically flat  solution of 5D Einstein-Maxwell gravity describing
a Saturn-like black object: a  rotating  black hole surrounded by a rotating dipole black ring. The solution is generated by combining
the  vacuum black Saturn solution
and  the  vacuum black ring solution with appropriately chosen parameters.

\section{Black Saturn with dipole ring }

A method for generating exact stationary solutions of 5D Einstein-Maxwell gravity was developed in \cite{Yazadjiev1}. This method
allows us to construct 5D Einstein-Maxwell solutions  by combining  two stationary solutions of the 5D vacuum Einstein equations.
Through this method, it was shown that  the dipole black ring can be constructed via the scheme
\begin{equation}
{\bf Rotating \,Black  \, Ring } \oplus {\bf Rotating \,Black  \, Ring} \longrightarrow {\bf Rotating\, Dipole  \,Black  \, Ring}
\end{equation}
where the black ring solutions are with appropriately chosen parameters. It is natural to expect that
 the generating scheme
\begin{equation}\label{GS2}
{\bf  Rotating \,Black  \, Ring } \oplus {\bf Black \,Saturn } \longrightarrow {\bf Black \,Saturn \,with \,Dipole \,Black\, Ring}
\end{equation}
will produce a black Saturn solution with a dipole black ring when the parameters of the solutions are appropriately chosen.

We present below the solution generated via the scheme (\ref{GS2}).  The spacetime geometry and the electromagnetic field are given by

\begin{eqnarray}
&&ds^2 = - {H_{2}\over H_{1} W} \left[dt + \left({\omega_{\psi}\over H_{2}} +q\right)d\psi \right]^2 +
k^2 {H_{1} P\over W} Y^3 \left(d\rho^2 + dz^2\right) + {H_{1}G_{\psi}\over H_{2}W } d\psi^2 + W^2 G_{\phi} d\phi^2, \nonumber  \\
&&F_{z\phi}= \sqrt{3} W^2 {G_{\phi}\over \rho} \partial_{\rho}{\cal V} ,\\
&&F_{\rho\phi}= - \sqrt{3} W^2 {G_{\phi}\over \rho} \partial_{z}{\cal V}. \nonumber
\end{eqnarray}
Here $k$ and $q$ are constants and the functions $W$, $Y$,  ${\cal V}$
are given by

\begin{eqnarray}
W = { {\tilde \mu}_{1} \left[\mu_{5}(\rho^2 + {\tilde \mu}_{1}\mu_{3} )^2(\rho^2 + {\tilde \mu}_{1}\mu_{4})^2 -
B^2_{1}\mu_{3}\mu_{4} ({\tilde \mu}_{1}-\mu_{5})^2\rho^4 \right] \over
\mu_{4}\left[\mu_{5}(\rho^2 + {\tilde \mu}_{1}\mu_{3} )^2(\rho^2 + {\tilde \mu}_{1}\mu_{4})^2 +
B^2_{1}{\tilde \mu}^2_{1}\mu_{3}\mu_{4} ({\tilde \mu}_{1}-\mu_{5})^2\rho^2 \right] },
\end{eqnarray}

\begin{eqnarray}
Y= {(\rho^2 + {\tilde \mu_{1}}\mu_{5})\left[\mu_{5}(\rho^2 + {\tilde \mu}_{1}\mu_{3} )^2(\rho^2 + {\tilde \mu}_{1}\mu_{4})^2 -
B^2_{1}\mu_{3}\mu_{4} ({\tilde \mu}_{1}-\mu_{5})^2\rho^4 \right] \over \mu_{5} (\rho^2 + {\tilde \mu}_{1}\mu_{3})
(\rho^2 + {\tilde \mu}_{1}\mu_{4})(\rho^2 + {\tilde \mu}^2_{1}) (\rho^2 +  \mu_{4}\mu_{5}) (\rho^2 + \mu_{3}\mu_{4})   },
\end{eqnarray}

\begin{eqnarray}
{\cal V} = B_{1}{\mu_{3}\mu_{4} \mu_{5}  ({\tilde \mu}_{1} -\mu_{5})  (\rho^2 + {\tilde \mu}^2_{1}) (\rho^2 + {\tilde \mu}_{1}\mu_{3})
(\rho^2 + {\tilde \mu}_{1}\mu_{4})  \over  {\tilde \mu}_{1} \left[\mu_{5}(\rho^2 + {\tilde \mu}_{1}\mu_{3} )^2(\rho^2 + {\tilde \mu}_{1}\mu_{4})^2 -
B^2_{1}\mu_{3}\mu_{4} ({\tilde \mu}_{1}-\mu_{5})^2\rho^4 \right] }.
\end{eqnarray}

The functions $P$, $G_{\psi}$, $G_{\phi}$, $\omega_{\psi}$, $H_{1}$ and $H_{2}$ are the metric functions of the vacuum black Saturn solution
\cite{ElvangFigueras} and
they are given by the following expressions

\begin{eqnarray}
  P =  (\mu_3\, \mu_4+ \rho^2)^2
      (\mu_1\, \mu_5+ \rho^2)
      (\mu_4\, \mu_5+ \rho^2) \, ,
\end{eqnarray}

\begin{eqnarray}
G_{\psi} = \frac{\mu_3\, \mu_5}{\mu_4},
\end{eqnarray}

\begin{eqnarray}
G_{\phi} = {\rho^2 \mu_{4}\over \mu_{3}\mu_{5}  },
\end{eqnarray}

\begin{eqnarray}
  \omega_\psi=
  2 \frac{
     c_{1}\, R_{1}\, \sqrt{M_0 M_1}
    -c_{2}\, R_{2}\, \sqrt{M_0 M_2}
    +c_{1}^2\,c_{2}\, R_{2}\, \sqrt{M_1 M_4}
    -c_{1}\,c_{2}^2\, R_{1}\, \sqrt{M_2 M_4}
  }
  {F \sqrt{G_{\phi}}} \, ,
\end{eqnarray}

\begin{eqnarray}
   H_{1} = F^{-1} \,
   \bigg[ M_0 + c_1^2 \, M_1 + c_2^2\,  M_2
   +  c_1\, c_2\, M_3 + c_1^2 c_2^2\, M_4 \bigg] \, ,
\end{eqnarray}
\begin{eqnarray}
    H_{2} = F^{-1} \,
   \frac{\mu_3}{\mu_4}\,
   \bigg[ M_0 \frac{\mu_1}{\mu_2}
   - c_1^2 \, M_1 \frac{\rho^2}{\mu_1\,\mu_2}
   - c_2^2\,  M_2 \frac{\mu_1\,\mu_2}{\rho^2}
   +  c_1\, c_2\, M_3
   + c_1^2 c_2^2\, M_4 \frac{\mu_2}{\mu_1} \bigg] \, ,
\end{eqnarray}
where
\begin{eqnarray}
 && M_{0} =\mu_2\, \mu_5^2 (\mu_1-\mu_3)^2 (\mu_2-\mu_4)^2
   (\rho^2+\mu_1\,\mu_2)^2(\rho^2+\mu_1\,\mu_4)^2
   (\rho^2+\mu_2\,\mu_3)^2 \, ,
\end{eqnarray}
\begin{eqnarray}
  &&M_{1} = \mu_1^2 \, \mu_2 \, \mu_3\, \mu_4 \, \mu_5 \, \rho^2\,
  (\mu_1-\mu_2)^2 (\mu_2-\mu_4)^2(\mu_1-\mu_5)^2
  (\rho^2+\mu_2\,\mu_3)^2  \, ,
\end{eqnarray}
\begin{eqnarray}
  &&M_{2} = \mu_2 \, \mu_3\, \mu_4 \, \mu_5 \, \rho^2\,
  (\mu_1-\mu_2)^2 (\mu_1-\mu_3)^2
  (\rho^2+\mu_1\,\mu_4)^2(\rho^2+\mu_2\, \mu_5)^2  \, ,
\end{eqnarray}
\begin{eqnarray}
  M_3 = &&2 \mu_1 \mu_2 \, \mu_3\, \mu_4 \, \mu_5 \,
  (\mu_1-\mu_3) (\mu_1-\mu_5)(\mu_2-\mu_4)
  (\rho^2+\mu_1^2)(\rho^2+\mu_2^2) \nonumber \\
  &&\times
  (\rho^2+\mu_1\,\mu_4)(\rho^2+\mu_2\, \mu_3)
  (\rho^2+\mu_2\, \mu_5)  \, ,
\end{eqnarray}
\begin{eqnarray}
  &&M_{4} = \mu_1^2 \, \mu_2\, \mu_3^2 \, \mu_4^2 \,
  (\mu_1-\mu_5)^2
  (\rho^2+\mu_1\,\mu_2)^2(\rho^2+\mu_2\, \mu_5)^2  \, ,
\end{eqnarray}
and
\begin{eqnarray}
  F &=& \mu_1\, \mu_5\,  (\mu_1-\mu_3)^2(\mu_2-\mu_4)^2
  (\rho^2+\mu_1\,\mu_3)
  (\rho^2+\mu_2\,\mu_3)
  (\rho^2+\mu_1\,\mu_4) \nonumber\\
  &&  \times
  (\rho^2+\mu_2\,\mu_4)
  (\rho^2+\mu_2\,\mu_5)
  (\rho^2+\mu_3\,\mu_5)
  \prod_{i=1}^5 (\rho^2+\mu_i^2) \, .
\end{eqnarray}
Finally we have
\begin{eqnarray}
  \mu_{i}=R_{i} - (z-a_{i}), \,\,\,  i=1,2,3,4,5
\end{eqnarray}
where
\begin{eqnarray}
 R_{i} = \sqrt{\rho^2 + (z-a_i)^2}
\end{eqnarray}
and
\begin{eqnarray}
{\tilde \mu}_{1} = \sqrt{\rho^2 + (z-b_{1})^2} - (z-b_{1}).
\end{eqnarray}
In all the above expressions $a_{i}$, $b_{1}$, $c_{1}$, $c_{2}$ and $B_{1}$ are (real) constants.
The presented solution possesses three Killing vectors, namely $\xi=\partial/\partial t$, $\zeta_{\psi}=\partial/\partial \psi$
and $\zeta_{\phi}=\partial/\partial \phi$.

Since $(c_{1},c_{2}) \to (-c_{1},-c_{2})$ and $B_{1} \to -B_{1}$ change the direction of rotation and the sign of the electromagnetic field
we can restrict ourselves to the case $c_{1}\ge 0$ and $B_{1}\ge 0$.
 Let us note some limiting cases of the presented solution. For $B_{1}=0$  and $b_{1}=a_{4}$ we obtain the black Saturn solution.
Setting further  $c_{1}=0$ and  $a_{1}=a_{5}=a_{4}$ we obtain the $\psi$-spinning Myers-Perry black hole.
Taking instead ($B_{1}\ne 0$) $c_{2}=0$ and  $a_{2}=a_{3}$ we obtain
the $S^{1}$ rotating dipole black ring. These limits will be discussed in more detail in  section 3.10.

\section{Analysis of the solution}

In order to analyze the solution we shall use the canonical procedure \cite{HAR} closely following the analysis of \cite{ElvangFigueras}.

As we have already mentioned, the generated solution describes regular black object when the solution parameters are chosen appropriately.
Here we will consider the following ordering of the parameters $a_{i}$ and $b_{1}$, namely

\begin{eqnarray}\label{Ordering}
a_{1}\le a_{5} \le a_{4} \le b_{1}< a_{3}\le a_{2}.
\end{eqnarray}
The  other bounds on the dipole parameter $b_{1}$ different form (\ref{Ordering}) lead to solutions which are singular and are not of physical interest.

In order to insure the positivity and regularity of the functions $W$, $Y$ and ${\cal V}$ we must impose the following
constraint

\begin{eqnarray}\label{RCON1}
B^2_{1} = 2{(a_{3}-b_{1})( b_{1}-a_{4})\over (b_{1}- a_{5})}.
\end{eqnarray}

Further, in order to remove the singularity of $H_{2}$  we must impose
the  constraint

\begin{eqnarray}\label{RCON2}
c^2_{1}= 2 {(a_{3}-a_{1})(a_{4}-a_{1})\over (a_{5}-a_{1}) }
\end{eqnarray}
just as for the black Saturn \cite{ElvangFigueras}.

Since the solution is invariant
under shifts in $z$-direction, its description in terms of $a_{i}$ and $b_{1}$ is redundant and it is convenient to introduce
new parameters. Following \cite{ElvangFigueras} we shall introduce one dimensionful parameter $L^2$ and four dimensionless parameters
$\kappa_{i}(i=1,2,3)$ and $\eta$ defined as follows

\begin{eqnarray}
&&L^2= a_{2}-a_{1}, \nonumber \\
&&\kappa_{i} = {a_{i+2} - a_{1}\over L^2 },\\
&&\eta = {b_{1}-a_{1}\over L^2 } \nonumber .
\end{eqnarray}

As a consequence of (\ref{Ordering}) we have

\begin{eqnarray}\label{Ordering1}
0\le \kappa_{3}\le \kappa_{2}\le \eta < \kappa_{1}\le 1.
\end{eqnarray}

In terms of the new parameters the conditions (\ref{RCON1}) and  (\ref{RCON2}) take the form

\begin{eqnarray}
&&B^2_{1} = 2L^2 {(\kappa_{1}-\eta)(\eta-\kappa_{2})\over (\eta - \kappa_{3}) },\\
&&c^2_{1} = 2L^2 {\kappa_{1}\kappa_{2}\over \kappa_{3} }.
\end{eqnarray}

It is also  convenient to introduce the dimensionless coordinate ${\bar z}$ and dimensionless parameter ${\bar c_{2}}$
given by \cite{ElvangFigueras}

\begin{eqnarray}
&&z=L^2 {\bar z} + a_{1},\\
&&{\bar c_{2}} = {c_{2}\over c_{1}(1-\kappa_{2}) }.
\end{eqnarray}

\subsection{Asymptotic behaviour of the solution}

In order to study the asymptotic behaviour of the solution we shall follow the standard way and we shall introduce
the asymptotic coordinates $r$ and $\theta$ defined by

\begin{eqnarray}
\rho = {1\over 2}r^2 \sin 2\theta ,\,\,\,  z = {1\over 2}r^2 \cos 2\theta.
\end{eqnarray}
Then in the asymptotic limit $r\to \infty$  we find

\begin{eqnarray}\label{ASFW}
W\approx 1 + 2{L^2\over r^2} (\eta -\kappa_{2}),
\end{eqnarray}

\begin{eqnarray}
Y\approx 1 + 2{L^2\over r^2} (\eta -\kappa_{2})\sin^2\theta ,
\end{eqnarray}

\begin{eqnarray}
{H_{1}\over H_{2}} \approx 1 + 2{L^2\over r^2} { \{\kappa_{3}(1-\kappa_{1} + \kappa_{2}) -2\kappa_{2}\kappa_{3}(\kappa_{1}-\kappa_{2}){\bar c}_{2}
+ \kappa_{2}\left[\kappa_{1} -\kappa_{2}\kappa_{3}(1+ \kappa_{1}-\kappa_{2}) \right]{\bar c}^2_{2} \} \over
\kappa_{3}\left(1 + \kappa_{2}{\bar c}_{2} \right)^2  } ,
\end{eqnarray}

\begin{eqnarray}
k^2 H_{1}P \approx {1\over r^2}k^2 \left(1 + \kappa_{2}{\bar c}_{2} \right)^2 ,
\end{eqnarray}

\begin{eqnarray}
&&{\omega_{\psi}\over H_{2}} \approx - L\sqrt{2\kappa_{1}\kappa_{2}\over \kappa_{3} } {{\bar c}_{2}\over \mid 1 + \kappa_{2}{\bar c}_{2}\mid}
 \\
&&+ {4L^3\over r^2 } {\sin^2\theta \over \kappa_{3}(1 + \kappa_{2}{\bar c}_{2})^3 }\sqrt{\kappa_{2}\over 2\kappa_{1}\kappa_{3} } \{\kappa^2_{3} -
\kappa_{3}\left[(\kappa_{1}-\kappa_{2}) (1-\kappa_{1} + \kappa_{3}) + \kappa_{2}(1-\kappa_{3})\right]{\bar c}_{2}  \nonumber \\
&&+
\kappa_{2}\kappa_{3}\left[(\kappa_{1}-\kappa_{2})(\kappa_{1}-\kappa_{3}) + \kappa_{1}(1 + \kappa_{1}- \kappa_{2}- \kappa_{3})\right]{\bar c}^2_{2}
\nonumber \\&&- \kappa_{1}\kappa_{2}\left[\kappa_{1} - \kappa_{2}\kappa_{3}(2+ \kappa_{1} - \kappa_{2} - \kappa_{3}) \right]{\bar c}^3_{2} \} ,
\nonumber
\end{eqnarray}

\begin{eqnarray}\label{ASFGPHI}
G_{\phi}\approx r^2\cos^2\theta ,\,\,\,\,\,\, G_{\psi} \approx r^2 \sin^2\theta .
\end{eqnarray}

The asymptotic behaviour of the function ${\cal V}$ is

\begin{eqnarray}
{\cal V}\approx 2  L^3 \sqrt{2(\kappa_{1}-\eta)(\eta - \kappa_{2})(\eta - \kappa_{3})} \,{\sin^2\theta \over r^2 }
\end{eqnarray}

Since we are interested in asymptotically flat solutions for which
$g_{t\psi}\to \sim\sin^2\theta/r^2$ and $g_{\rho\rho}\to 1/ r^2$ we chose the constants $q$ and $k$ to be

\begin{eqnarray}
q = L\sqrt{2\kappa_{1}\kappa_{2}\over \kappa_{3}} {{\bar c}_{2}\over 1 + \kappa_{2}{\bar c}_{2} },
\end{eqnarray}
\begin{eqnarray}
k = \mid 1 + \kappa_{2} {\bar c}_{2}\mid^{-1} ,
\end{eqnarray}
for $\kappa_{2} {\bar c}_{2}\ne -1$. The case  $\kappa_{2} {\bar c}_{2}= -1$ is singular and will not be considered  here.
With this choice for $q$ and $k$ the asymptotic metric takes the form

\begin{eqnarray}
ds^2 \approx -dt^2 + dr^2 + r^2d\theta^2 + r^2\sin^2\theta d\psi^2 + r^2\cos^2\theta d\phi^2 .
\end{eqnarray}
In what follows we shall show that the periodicities of the angles $\psi$ and $\phi$ can be chosen to be canonical ones
\begin{eqnarray}\label{ASF}
\Delta \psi = \Delta \phi = 2\pi
\end{eqnarray}
and as a consequence  we find that the spacetime is asymptotically flat.

\subsection{Rod structure and balance conditions }

As we have already mentioned the ordering (\ref{Ordering}) and the condition (\ref{RCON1}) insure the positivity and
the regularity of the functions $W$ and $Y$. Therefore the rod structure \cite{HAR} of the solution presented here is the same
as the rod structure of the black Saturn. With  condition (\ref{RCON2}) imposed the rod structure is the following.
We have

$\ast$ Semi-infinite rod ${\bar z} \in [-\infty, \kappa_{3}]$ and finite rod ${\bar z} \in [\kappa_{2}, \kappa_{1}]$
in direction $(0,1,0)$ which are sources of the $\phi\phi$-part of the metric

$\ast$ The source of the $\psi\psi$-part of the metric is the semi-infinite rod ${\bar z} \in [1,\infty]$ in direction
$(0,0,1)$.

$\ast$  Two finite rods ${\bar z} \in [\kappa_{3}, \kappa_{2}]$ and ${\bar z} \in [\kappa_{1}, 1]$ in directions
$(1,0,\Omega^{DBR}_{\psi})$ and $(1,0,\Omega^{BH}_{\psi})$. They correspond to the location the dipole black ring horizon
and the location of the black hole, respectively. $\Omega^{DBR}_{\psi}$ and $\Omega^{BH}_{\psi}$
are the angular velocities of the horizons.

In order to cure the conical singularities at the location of a rod the coordinates $\psi$ and $\phi$ should have periods

\begin{eqnarray}\label{Periods}
\Delta \psi = 2\pi \lim_{\rho\to 0} \sqrt{\rho^2 g_{\rho\rho}\over g_{\psi\psi}}, \,\,\,\,\,
\Delta \phi = 2\pi \lim_{\rho\to 0} \sqrt{\rho^2 g_{\rho\rho}\over g_{\phi\phi}} .
\end{eqnarray}

Let us first consider the rods ${\bar z} \in [-\infty, \kappa_{3}]$ and ${\bar z} \in [1,\infty]$. Then the regularity conditions\footnote{Let us
note again that we keep imposing the conditions (\ref{RCON1}) and (\ref{RCON2}) .}
(\ref{Periods}) fix the periods of $\phi$ and $\psi$ to be $\Delta\phi=2\pi$ and $\Delta\psi= 2\pi$. These periods insure asymptotic flatness
of the metric as it was discussed in (\ref{ASF}).

Concerning the regularity on the finite rod ${\bar z} \in [\kappa_{2}, \kappa_{1}]$ we obtain

\begin{eqnarray}\label{Periodphi}
\Delta \phi= 2\pi {(\kappa_{1}-\eta)^{3/2}\over  \mid 1 + \kappa_{2}{\bar c}_{2} \mid
\sqrt{\kappa_{1}(1-\kappa_{2})(1-\kappa_{3}) (\kappa_{1}-\kappa_{2})(\kappa_{1}-\kappa_{3}) }  }.
\end{eqnarray}

In order to  find the equilibrium (the balancing) condition for the black Saturn with dipole black ring we have to impose that
the r.h.s. of (\ref{Periodphi}) be equal  to $2\pi$. Solving then for ${\bar c}_{2}$ we find

\begin{eqnarray}\label{Balancecon}
{\bar c}_{2} = {1\over \kappa_{2}} \left[\varepsilon {(\kappa_{1}-\eta)^{3/2}\over
 \sqrt{\kappa_{1}(1-\kappa_{2})(1-\kappa_{3}) (\kappa_{1}-\kappa_{2})(\kappa_{1}-\kappa_{3}) } } -1 \right]
\end{eqnarray}
where $\varepsilon=1$ for ${\bar c}_{2}>-\kappa_{2}^{-1}$ and $\varepsilon=-1$ for ${\bar c}_{2}<-\kappa_{2}^{-1}$.
As in the case of black Saturn, we have two separate sectors of the balance condition for the black Saturn with a dipole
black ring.

Let us also note that the electromagnetic field is regular on the rods considered here.

\subsection{Black hole horizon}

The black hole horizon is located at $\rho=0$ for ${\bar z} \in [\kappa_{1}, 1]$. The induced metric on the spacial cross section of the horizon
is given by

\begin{eqnarray}
  ds^2_{BH} =
  \frac{2L^2 ({\bar z}-\kappa_{1})({\bar z}-\kappa_{3})}{({\bar z}-\kappa_2)} w^2({\bar z}) d\phi^2
  +  L^2 s^2_{BH} \, {g({\bar z})\over w({\bar z})} (1-{\bar z})\, d\psi^2 \nonumber \\
  + \frac{L^2\, ({\bar z}-\kappa_2)\, d{\bar z}^2}
         {(1-{\bar z})({\bar z}-\kappa_{1})({\bar z}-\kappa_{3})g({\bar z})w({\bar z}) } \, .~~~
\end{eqnarray}

Here  the constant $s_{BH}$ and the function $g({\bar z})$ are formally the same\footnote{Let us note however that
the balance condition  (\ref{Balancecon}) is different  i.e. the parameter ${\bar c}_{2}$ is different. } as for the black Saturn solution \cite{ElvangFigueras} and are given by
\begin{eqnarray}
  s_{BH}=
  \frac{\kappa_{3}(1-\kappa_{1})
  +\kappa_{1} \kappa_{2} (1-\kappa_{2}) (1-\kappa_{3}) {\bar c}^2_{2}}
    {\kappa_{3} \sqrt{(1-\kappa_{1}) (1-\kappa_{2}) (1-\kappa_{3})}
      \,\big[ 1+ \kappa_{2} {\bar c}_2 \big]^2} \, ,
\end{eqnarray}

\begin{eqnarray}
  \nonumber
  g({\bar z}) &=& 2 \kappa_{1} \kappa_{3} (1-\kappa_{1}) (1-\kappa_{2}) (1-\kappa_{3})  ({\bar z}-\kappa_{2}) \\[2mm]
    &&\nonumber \times
    \big[ 1+ \kappa_{2} {\bar c}_2 \big]^2
    \bigg[
      (1-\kappa_{1})^2 \kappa_{3}
      \Big[ \kappa_1 ({\bar z} - \kappa_{2})
         - \kappa_{3} \Big( \kappa_{1}- \kappa_{2}(1-{\bar z})^2 - \kappa_{1} \kappa_{2}(2-{\bar z})\Big)
      \Big]  \\[2mm]
    &&\nonumber \hspace{1.3cm}
    + 2 \kappa_{1} \kappa_{2} \kappa_{3}   (1-\kappa_{1}) (1-\kappa_{2}) (1-\kappa_3)
      (1-{\bar z}) ({\bar z} - \kappa_{1}) \, {\bar c}_2 \\[2mm]
    && \hspace{1.3cm}
    + \kappa^2_{1} \kappa_{2} (1-\kappa_{2})^2 (1-\kappa_{3})^2 \, {\bar z}\, ({\bar z} - \kappa_{1})\, {\bar c}_2^2
    \bigg]^{-1} \, .
\end{eqnarray}

The function $w({\bar z})$ is defined by

\begin{eqnarray}
w({\bar z}) = {{\bar z} - \kappa_{2}\over {\bar z} -\eta }.
\end{eqnarray}

The functions $g({\bar z})$ and $w({\bar z})$ are positive for $\kappa_{1}\le {\bar z}\le 1$ and $s_{BH}>0$.

The topology of the horizon is that of $S^3$. Metrically, however, the horizon is distorted $S^{3}$. This can be seen
explicitly by computing the scalar curvature of the horizon which is not constant contrary to the case of the round $S^3$. The distortion
is caused by the rotation of the black hole itself and the gravitational attraction of the dipole black ring.

The area of the spacial cross section  black hole  horizon ca be found  by straightforward calculation and the result
is

\begin{eqnarray}
{\cal A}_{BH}= 4\pi^2 L^3 \sqrt{{2(1-\kappa_{1})^3 \over (1-\kappa_{2})(1-\kappa_{3}) }}
{\kappa_{3}(1-\kappa_{1}) + \kappa_{1}\kappa_{2}(1-\kappa_{2})(1-\kappa_{3}){\bar c}^2_{2}\over \kappa_{3}(1-\kappa_{1})
(1 + \kappa_{2}{\bar c}_{2})^2  }.
\end{eqnarray}

The other horizon quantities which are of interest are the angular velocity and the temperature and they are given by

\begin{eqnarray}
\Omega^{BH}_{\psi}= {1\over L} \sqrt{\kappa_{2}\kappa_{3}\over 2\kappa_{1}} {\kappa_{3}(1-\kappa_{1})
- \kappa_{1}(1-\kappa_{2}) (1-\kappa_{3}){\bar c}_{2}\over \kappa_{3}(1-\kappa_{1}) +
\kappa_{1}\kappa_{2}(1-\kappa_{2}) (1-\kappa_{3}){\bar c}^2_{2} } (1 + \kappa_{2}{\bar c}_{2}),
\end{eqnarray}

\begin{eqnarray}
T_{BH}= {1\over 2\pi L} \sqrt{(1-\kappa_{2})(1-\kappa_{3})\over 2(1-\kappa_{1})}  { \kappa_{3}(1-\kappa_{1})
(1 + \kappa_{2}{\bar c}_{2})^2  \over \kappa_{3}(1-\kappa_{1}) + \kappa_{1}\kappa_{2}(1-\kappa_{2})(1-\kappa_{3}){\bar c}^2_{2} }.
\end{eqnarray}
 Let us note that although the horizon area, temperature and the angular velocity of the black hole in the black Saturn
with dipole ring look at first sight the same as the corresponding quantities for the black hole in the black Saturn solution,
they are in fact different since the balance condition (\ref{Balancecon}) (i.e. the parameter ${\bar c}_{2}$) is different.

The investigation of the behaviour of the electromagnetic field shows that it  is regular on the black hole horizon.

\subsection{Dipole black ring horizon}

The dipole black ring horizon is located at $\rho=0$ for ${\bar z}\in [\kappa_{3},\kappa_{2}]$. The induced metric on the
spacial cross section of the horizon is

\begin{eqnarray}
  ds^2_{DBR} = \frac{2L^2 (\kappa_{2}-{\bar z})({\bar z}-\kappa_{3})}{(\kappa_{1}-{\bar z})} w^2({\bar z}) d\phi^2
  +  L^2 s_{DBR}^2\, {f({\bar z})\over w({\bar z})} (\kappa_{1}-{\bar z})\, d\psi^2 \nonumber \\
  + \frac{L^2 y^3_{DBR}d{\bar z}^2}{ (\kappa_{2}-{\bar z})({\bar z}-\kappa_{3})f({\bar z}) w({\bar z})} \, .
\end{eqnarray}
Here the constant $s_{DBR}$ and the function $f({\bar z})$ are formally the same as for the black Saturn solution \cite{ElvangFigueras} and are given by
\begin{eqnarray}
  s_{DBR}= \sqrt{\frac{\kappa_{2}(\kappa_{2}-\kappa_{3})}{\kappa_{1} (\kappa_{1}-\kappa_{3}) (1-\kappa_{3})}}
  \,\frac{\left[ \kappa_{3}- \kappa_{3}(\kappa_{1}-\kappa_{2}) {\bar c}_2 +\kappa_{1} \kappa_{2} (1-\kappa_{3})
    {\bar c}_2^2 \right]}
    {\kappa_{3}(1+ \kappa_{2} {\bar c}_2 )^2} \, ,
\end{eqnarray}

\begin{eqnarray}
  f({\bar z}) &=& 2\kappa_{1}\kappa_{3}(\kappa_{1}-\kappa_{3})(1-\kappa_{3})(1-{\bar z})\nonumber\\
    &\;&\times
    \big[ 1+ \kappa_{2} {\bar c}_2 \big]^2 (\kappa_{2}-\kappa_{3})^{-1}\bigg[
    \kappa_{3}\Big[\kappa_{2}(\kappa_{1}-{\bar z})+\kappa_{3}\Big(\kappa_{2}\big(1-\kappa_{1}(2-{\bar z})\big)-\kappa_{1}(1-{\bar z})^2\Big)\Big]
\nonumber\\
    &\;&+2\kappa_{1}\kappa_{2}\kappa_{3}(1-\kappa_{3})(1-{\bar z})(\kappa_{2}-{\bar z}){\bar c}_2\nonumber\\
    &\;&+\kappa_{1}\kappa_{2}^2(1-\kappa_{3})^2{\bar z}(\kappa_{2}-{\bar z}){\bar c}_2^2
    \bigg]^{-1}\;.
\end{eqnarray}
The function $w({\bar z})$ and the constant  $y_{DBR}$ are defined by

\begin{eqnarray}
w({\bar z}) = {(\eta - {\bar z}) (\kappa_{1} -{\bar z})  \over (\kappa_{1} - {\bar z}) (\kappa_{2} - {\bar z})
+ {(\kappa_{1}-\eta)(\eta- \kappa_{2})\over (\eta - \kappa_{3})} ({\bar z}- \kappa_{3}) } ,
\end{eqnarray}

\begin{eqnarray}
y_{DBR}=  {\eta - \kappa_{3}\over \kappa_{2}- \kappa_{3}} .
\end{eqnarray}

The functions $f({\bar z})$ and $w({\bar z})$ are positive for $\kappa_{3}\le {\bar z}\le \kappa_{2}$ and $s_{DBR}>0$, $y_{DBR}>0$.
The topology of the horizon is $S^2\times S^1$ where $S^{1}$ is parameterized by the angular coordinate $\psi$ and has radius depending on ${\bar z}$.
The  two-sphere is
parameterized by the coordinates $({\bar z},\phi)$ and is metrically distorted.

The horizon area is found by straightforward integration and the result is

\begin{eqnarray}
{\cal A}_{DBR} = 4\pi^2 L^3 \sqrt{2\kappa_{2}(\eta- \kappa_{3})^3\over \kappa_{1}(\kappa_{1}-\kappa_{3})(1-\kappa_{3}) }
{\kappa_{3} - \kappa_{3}(\kappa_{1}- \kappa_{2}){\bar c}_{2} + \kappa_{1}\kappa_{2}(1-\kappa_{3}){\bar c}^2_{2}\over
\kappa_{3}(1 + \kappa_{2}{\bar c}_{2})^2 } .
\end{eqnarray}

The angular velocity and the temperature of the horizon are given by

\begin{eqnarray}
\Omega^{DBR}_{\psi} = {1\over L} \sqrt{\kappa_{1}\kappa_{3}\over 2\kappa_{2} } {\kappa_{3} - \kappa_{2}(1-\kappa_{3}){\bar c}_{2} \over
\kappa_{3} - \kappa_{3}(\kappa_{1}-\kappa_{2}){\bar c}_{2} + \kappa_{1}\kappa_{2}(1 -\kappa_{3}){\bar c}^2_{2} }(1 + \kappa_{2}{\bar c}_{2}),
\end{eqnarray}

\begin{eqnarray}
T_{DBR} = {1\over 2\pi L } \sqrt{ \kappa_{1}(1-\kappa_{3})(\kappa_{1}-\kappa_{3})(\kappa_{2}-\kappa_{3})^2 \over 2\kappa_{2}(\eta - \kappa_{3})^3 }
{\kappa_{3}(1 + \kappa_{2}{\bar c}_{2})^2 \over \kappa_{3} - \kappa_{3}(\kappa_{1}-\kappa_{2}){\bar c}_{2}
+ \kappa_{1}\kappa_{2} (1-\kappa_{3}){\bar c}^2_{2} }.
\end{eqnarray}

The analysis shows that the electromagnetic field is regular on the black ring horizon.

\subsection{ADM mass and angular momentum of the black Saturn with dipole black ring}

The ADM mass and angular momentum can be found from the asymptotic form of the metric (\ref{ASFW})-(\ref{ASFGPHI}) and the result is

\begin{eqnarray}
&&M_{ADM} = {3\pi\over 4} L^2(\eta - \kappa_{2}) \\
&&+ {3\pi\over 4}L^2 { \{\kappa_{3}(1-\kappa_{1} + \kappa_{2}) -2\kappa_{2}\kappa_{3}(\kappa_{1}-\kappa_{2}){\bar c}_{2}
+ \kappa_{2}\left[\kappa_{1} -\kappa_{2}\kappa_{3}(1+ \kappa_{1}-\kappa_{2}) \right]{\bar c}^2_{2} \} \over
\kappa_{3}\left(1 + \kappa_{2}{\bar c}_{2} \right)^2  }  \nonumber ,
\end{eqnarray}

\begin{eqnarray}
&&J_{ADM} = {\pi L^3 \over \kappa_{3}(1 + \kappa_{2}{\bar c}_{2})^3 }\sqrt{\kappa_{2}\over 2\kappa_{1}\kappa_{3} } \{\kappa^2_{3} -
\kappa_{3}\left[(\kappa_{1}-\kappa_{2}) (1-\kappa_{1} + \kappa_{3}) + \kappa_{2}(1-\kappa_{3})\right]{\bar c}_{2}  \nonumber  \\
&&+
\kappa_{2}\kappa_{3}\left[(\kappa_{1}-\kappa_{2})(\kappa_{1}-\kappa_{3}) + \kappa_{1}(1 + \kappa_{1}- \kappa_{2}- \kappa_{3})\right]{\bar c}^2_{2}
\\&&- \kappa_{1}\kappa_{2}\left[\kappa_{1} - \kappa_{2}\kappa_{3}(2+ \kappa_{1} - \kappa_{2} - \kappa_{3}) \right]{\bar c}^3_{2} \} \nonumber
\end{eqnarray}.

Let us note that the ADM mass is positive as a consequence of the ordering (\ref{Ordering1}) .

\subsection{Komar masses and angular momenta}

The definition of the Komar mass and angular momentum is well known, namely

\begin{eqnarray}
M_{Komar} = {3\over 32\pi} \int_{\partial\Sigma} \star d{\tilde \xi} ,\,\,\, \,
J_{Komar} = {1\over 16\pi} \int_{\partial\Sigma}\star d{\tilde \zeta}_{\psi}
\end{eqnarray}
where ${\tilde \xi}$ and ${\tilde \zeta}_{\psi}$ are $1$-forms dual to the timelike Killing vector $\xi$ and the spacelike Killing vector
$\zeta_{\psi}$. Here $\partial\Sigma$ is a boundary of any spacelike hypersurface $\Sigma$. From a physical point of view the Komar integrals
measure the mass and angular momentum contained in $\partial\Sigma$. When $\partial\Sigma$ is a three-sphere at infinity the Komar
integrals coincide with the ADM mass and angular momentum of an asymptotically flat spacetime. When dealing with multi-black objects
configurations the Komar integrals evaluated on the horizon cross sections  are of special interest since they give the
intrinsic mass and angular momenta of the black objects.  These intrinsic quantities for the black objects in our solution are
the following

\begin{eqnarray}
M^{BH}_{Komar} = {3\pi L^2 \over 4} {\kappa_{3}(1-\kappa_{1}) + \kappa_{1}\kappa_{2}(1-\kappa_{2})(1-\kappa_{3}){\bar c}_{2}^2\over
\kappa_{3}(1+ \kappa_{2}{\bar c}_{2})  },
\end{eqnarray}

\begin{eqnarray}
M^{DBR}_{Komar}=  {3\pi L^2 \over 4} {\kappa_{2}[1 - (1-\kappa_{2}){\bar c}_{2}][\kappa_{3} - \kappa_{3}(\kappa_{1}-\kappa_{2}){\bar c}_{2}
+ \kappa_{1}\kappa_{2}(1-\kappa_{3}){\bar c}^2_{2}] \over \kappa_{3}(1+ \kappa_{2}{\bar c}_{2})^2 },
\end{eqnarray}

\begin{eqnarray}
J^{BH}_{Komar} = - \pi L^3 {\bar c}_{2}\sqrt{\kappa_{1}\kappa_{2}\over 2\kappa_{3} } {\kappa_{3}(1-\kappa_{1}) +
\kappa_{1}\kappa_{2}(1-\kappa_{2})(1-\kappa_{3}){\bar c}^2_{2}\over \kappa_{3}(1+ \kappa_{2}{\bar c}_{2})^2},
\end{eqnarray}

\begin{eqnarray}
&&J^{DBR}_{Komar} =
\pi L^3  \sqrt{\kappa_{2}\over 2\kappa_{1}\kappa_{3} }\\&&\times {[\kappa_{3} - \kappa_{2}(\kappa_{1}-\kappa_{3}){\bar c}_{2}
+ \kappa_{1}\kappa_{2}(1-\kappa_{2}){\bar c}^2_{2}] [\kappa_{3} - \kappa_{3}(\kappa_{1}-\kappa_{2}){\bar c}_{2}
+ \kappa_{1}\kappa_{2}(1-\kappa_{3}){\bar c}^2_{2}]\over \kappa_{3}(1 + \kappa_{2}{\bar c}_{2})^3 } \nonumber.
\end{eqnarray}

Usually one thinks that the Komar mass of every black object in a selfgravitating configuration should be positive.
However, when the objects are under strong gravitational interaction the Komar masses can become negative for
such tightly bound configurations. Numerical solutions of black holes with negative Komar mass but with positive ADM mass were
presented in \cite{Kunz1},\cite{Kunz2} for the 5D Einstein-Maxwell gravity with a Chern-Simons term. For our solution the situation
is the same as for the vacuum black Saturn solution \cite{ElvangFigueras}. The balance condition (\ref{Balancecon}) with $\varepsilon=1$
leads to the inequality $-\kappa^{-1}_{2}<{\bar c}_{2}<(1-\kappa_{2})^{-1}$ which guarantees that $M^{BH}_{Komar}>0$ and $M^{DBR}_{Komar}>0$.
The balance condition (\ref{Balancecon}) with $\varepsilon=-1$ leads to the inequality ${\bar c}_{2}<-\kappa_{2}^{-1}$ and as a consequence
we have  $M^{BH}_{Komar}<0$ and $M^{DBR}_{Komar}>0$.

\subsection{Dipole charge and potential}

The dipole charge is defined as

\begin{eqnarray}
{\cal Q} = {1\over 4\pi} \oint_{S^2} F
\end{eqnarray}
where $S^2$ is a sphere on the black ring horizon. For our solution we find

\begin{eqnarray}
{\cal Q} = - L \sqrt{3(\eta-\kappa_{2})(\eta-\kappa_{3}) \over 2(\kappa_{1} -\eta) }.
\end{eqnarray}

Finding explicitly the potential $B$ ($H=dB$) of the  dual form $H=*F$ seams to be formidable task at least in the canonical coordinates $(\rho,z)$.
That is why the direct computation of the dipole potential $\Phi$ is not possible. In order to find $\Phi$ we shall proceed in the following
way. First it is clear that $\Phi$ is proportional to $\sqrt{3}B_{1}$ i.e.

\begin{eqnarray}
\Phi = \Gamma \sqrt{3} B_{1}
\end{eqnarray}
where the dimensionless constant $\Gamma$  is a function of the solution parameters. What is important is that this constant
does not depend on the parameters
$a_{2}$ and  $c_{2}$. Therefore $\Gamma$ is the same for the dipole black ring solution which is obtained in the limit $c_{2}=0$ and $a_{2}=a_{3}$.
Hence  we find $\Gamma= - {\pi\over 2} $ and therefore we have

\begin{eqnarray}
\Phi = -{\sqrt{3}\pi\over 2} B_{1}= -\pi L\sqrt{{3(\kappa_{1}-\eta)(\eta-\kappa_{2})\over 2(\eta-\kappa_{3}) }}.
\end{eqnarray}

\subsection{Smarr relations}

Straightforward calculations show that the following Smarr-like relations are satisfied

\begin{eqnarray}
M^{BH}_{Komar} = {3\over 2} \left[T^{BH}{{\cal A}_{BH}\over 4} + J^{BH}_{Komar}\Omega^{BH}_{\psi}\right],
\end{eqnarray}

\begin{eqnarray}
M^{DBR}_{Komar} = {3\over 2}\left[ T^{DBR}{{\cal A}_{DBR}\over 4} + J^{DBR}_{Komar}\Omega^{DBR}_{\psi}\right],
\end{eqnarray}

\begin{eqnarray}
M_{ADM} =  M^{BH}_{Komar} + M^{DBR}_{Komar} + {1\over 2} {\cal Q} \Phi .
\end{eqnarray}

The term ${1\over 2}{\cal Q}\Phi$ can be interpreted as the energy of the electromagnetic "hair" of the dipole black ring.

\subsection{Ergosurfaces and closed timelike curves}

As a consequence of the way the solution was generated and the positiveness of $W$ the existence/noexistence of ergosurfaces and closed timelike curves
depends only on the seed vacuum black Saturn solution i.e. the dipole solution inherits
the existence/nonexistence of ergosurfaces and closed timelike curves from the vacuum black Saturn solution. Our preliminary
investigations show that there are ergosurfaces for the  black hole and the black ring. However, the solution is too
involved in the canonical coordinates and the explicit description is not possible. Concerning the possible existence
of closed timelike curves we should say that no sign for their existence is seen. The same conclusion is also reached in
\cite{ElvangFigueras}. The problem however remains open.

\subsection{Limits of the solution}

The limits of the balance black Saturn with dipole black ring are more or less clear from the way the solution was generated.
By removing the black hole from the configuration we obtain the dipole black ring \cite{EMP},\cite{Yazadjiev1}. The formal procedure
is as follows. First we  set the angular momentum of the black hole to zero by taking ${\bar c}_{2}=0$. Then the black hole is removed
by setting $\kappa_{1}=1$.

Removing the dipole black ring from the configuration we obtain a Myers-Perry black hole with one angular momentum. This is achieved as follows.
First we set $\eta=\kappa_{2}$ which eliminates the electromagnetic field from the black ring. Then we take the limit
$\kappa_{3}\to \kappa_{2}$ and finally set $\kappa_{2}=0$.

It is tempting to consider limit in which the black ring is removed but the electromagnetic field is preserved i.e. to repeat the above
procedure without setting $\eta=\kappa_{2}$ expecting that the result is a dipole black hole. Unfortunately the described limit is
singular as the analysis shows. One can  show that there is no regular limit of  a merger of the black hole and the dipole black ring.

\subsection{Non-uniqueness}

The balanced solution  depends on four dimensionless parameters
$(\kappa_{1},\kappa_{2},\kappa_{3},\eta)$ and one dimensionful parameter $L$. Two of the parameters can be
fixed by fixing the mass and angular momentum of the configuration. Therefore the solution exhibits 3-fold continuous non-uniqueness.

\section{Discussions}

In this paper we  presented a new asymptotically flat  solution of 5D Einstein-Maxwell gravity describing a Saturn-like black configuration:
a  rotating  black hole surrounded by a rotating dipole black ring. The solution was generated by combining the  vacuum black Saturn solution
and  the  vacuum black ring solution with appropriately chosen parameters along the lines of the method developed in \cite{Yazadjiev1}.
Some basic properties of the solution were analyzed and the basic quantities were calculated. It is interesting to see in detail how
the presence of the dipole charge affects the physics of the black Saturn. This however  is  very difficult and requires numerical methods
since the dipole black Saturn depends on many parameters in nontrivial manner. The physical properties of the dipole black Saturn are currently
under numerical investigation and the results will be presented elsewhere. Preliminary results show that, more or less,
 many of the physical properties of the dipole solution are similar to those of the vacuum black Saturn solution which are thoroughly enough
discussed in \cite{ElvangFigueras}. In particular, the dipole black Saturn exhibits effects like rotational frame-dragging and
countering frame-dragging. Respectively, the differences  between the dipole black Saturn and the vacuum black Saturn are inherited from
the differences of the dipole black ring and the  vacuum black ring \cite{EMP}. In particular, the  dipole charge increases the self-interaction   of the ring
and larger angular momentum is needed to balance the ring in comparison with the vacuum black Saturn solution. Another important point is that
the dipole black Saturn is expected to be stable near the extremal non-BPS limit \cite{EEF}.

Lets us finish with prospects for future investigations.
The question whether there is a more general dipole Saturn solution
than the one presented here remains open. If such a solution exists, most probably it has to be generated  via the scheme

\begin{equation}
{\bf Black \,Saturn }\oplus  {\bf Black \,Saturn }   \longrightarrow {\bf New \,Dipole \,Black \,Saturn\,?}
\end{equation}
provided that the potential singularities can be removed by an appropriate choice of the solution parameters.
Since the  formal operation denoted by $\bigoplus$ is not commutative it is also interesting to consider the
 generating scheme
\begin{equation}
{\bf Black \,Saturn }\oplus  {\bf Rotating \,Black  \, Ring }   \longrightarrow {\bf New \,Dipole \,Black \,Rings \, ?}
\end{equation}
which will generate new dipole black ring solution provided the potential singularities can be removed.

Our solution can be extended to a solution of the 5D Einstein-Maxwell-dilaton gravity with an arbitrary dilaton coupling parameter
via the solution generating methods developed in \cite{Yazadjiev2}, \cite{Yazadjiev3}. Moreover these dipole solutions can be uplifted
to supergravity solutions.

Finally the methods of \cite{Yazadjiev1}, \cite{Yazadjiev2} and  \cite{Yazadjiev3} can be applied to the  vacuum solution
of \cite{IgichiMishima} describing two (or more) rotating back rings at equilibrium  in order generate  configurations with
 two (or more) dipole black rings.

\section*{Acknowledgements}
The author would like to thank the Alexander von Humboldt Foundation for a stipend, and
the Institut f\" ur Theoretische Physik G\" ottingen for its kind hospitality. The partial support by the
Bulgarian National Science Fund under Grant MUF04/05 (MU 408) and VUF-201/06  is also acknowledged.


\begin{thebibliography}{tbds}









\bibitem{ER2} R.~Emparan~and~H.~Reall, ~Phys.~Rev.~Lett.~{\bf 88},~101101~(2002); hep-th/0110260
\bibitem{Pomeransky} A.~Pomeransky~and~A.~Sen'kov, hep-th/0612005
\bibitem{EMP} R.~Emparan,~JHEP~{\bf 0403},~064~(2004); hep-th/0402149





\bibitem{ElvangFigueras} H.~Elvang~and~P.Figueras, {\it Black Saturn},
hep-th/0701035

\bibitem{IgichiMishima} H.~Iguchi~and~T.~Mishima,~Phys.~Rev.~{\bf D75}, 064018 (2007); hep-th/0701043

\bibitem{Gauntlett1} J.~Gauntlett~and~J.~Gutowski,~Phys.~Rev.~{\bf D71}, 025013 (2005); hep-th/0408010


\bibitem{Gauntlett2} J.~Gauntlett~and~J.~Gutowski,~Phys.~Rev.~{\bf D71}, 045002 (2005); hep-th/0408122


\bibitem{Yazadjiev1} S.~Yazadjiev,~Phys.~Rev.~{\bf D73}, 104007 (2006); hep-th/0602116

\bibitem{HAR} T.~Harmark,~Phys.~Rev.~{\bf D70},~124002~(2004); hep-th/0408141

\bibitem{Kunz1} J.~Kunz~and~F.~Navarro-Lerida, Phys. Lett. {\bf B643}, 55 (2006); hep-th/0610036
\bibitem{Kunz2} J.~Kunz~and~F.~Navarro-Lerida, Mod.~Phys.~Lett.~{\bf A21}, 2621 (2006); hep-th/0610075

\bibitem{EEF} H.~Elvang,~R.~Emparan~and~P.~Figueras, hep-th/0702111


\bibitem{Yazadjiev2} S.~Yazadjiev,~JHEP~{\bf 0607}, 036 (2006);  hep-th/0604140

\bibitem{Yazadjiev3} S.~Yazadjiev,~Gen.~Rel.~Grav.~{\bf 39}, 601 (2007); hep-th/0607101


\end{thebibliography}
\end{document}